\documentclass[10pt,prl,aps,twocolumn,showpacs,floatfix]{revtex4}
\usepackage{graphicx}

\usepackage{amssymb}

\begin{document}

\title{Ground states of a frustrated quantum spin chain with long-range interactions}

\author{Anders W. Sandvik}
\affiliation{Department of Physics, Boston University, 590 Commonwealth Avenue, Boston, Massachusetts 02215}

\begin{abstract}
The ground state of a spin-$1/2$ Heisenberg chain with both frustration and long-range interactions is studied using Lanczos exact 
diagonalization. The evolution of the well known dimerization transition of the system with short-range frustrated interactions 
(the J$_{\rm 1}$-J$_{\rm 2}$ chain) is investigated in the presence of additional unfrustrated interactions decaying with distance
as $1/r^\alpha$. It is shown that the continuous (infinite-order) dimerization transition develops into a first-order transition between 
a long-range ordered antiferromagnetic state and a state with coexisting dimerization and critical spin correlations at wave-number
$k=\pi/2$. The relevance of the model to real systems is discussed.
\end{abstract}

\date{\today}


\maketitle

One-dimensional spin systems have played an important role in quantum many-body physics since the early days of quantum 
mechanics \cite{bethe,hulthen}. Several different types of ordered and disordered ground states can be realized, depending on the 
individual spin magnitude $S$ and the form of the spin-spin interactions \cite{majumdar,haldane1,affleck1,affleck2}. For $S=1/2$, 
the prototypical Heisenberg chain with antiferromagnetic nearest-neighbor interactions (coupling constant $J_1>0$) has a quasi-ordered 
(critical)  ground state, with spin correlations decaying with distance $r$ as $\sqrt{{\rm ln}(r)}/r$ \cite{affleck2}. 
Including a next-nearest-neighbor coupling $J_2>0$ (the J$_{\rm 1}$-J$_{\rm 2}$ chain \cite{majumdar}) leads to a quantum phase transition into a 
doubly-degenerate dimerized state (a valence-bond-solid; VBS) at coupling ratio $g=J_2/J_1 \approx 0.2411$. In the effective field theory
for the $S=1/2$ chain \cite{affleck1}, the VBS transition is related to a sign change of a marginal operator. It has been investigated in 
great detail numerically, using, e.g., exact diagonalization \cite{nomura,eggert} and the density-matrix renormalization-group (DMRG) 
method \cite{bursill}. 

While long-range spin ordering is rigorously ruled out in one-dimensional systems with finite-range rotationally invariant interactions, 
long-range interactions make magnetic order possible at zero temperature. The transition between a long-range ordered antiferromagnet (AFM) and 
the quasi-long-range ordered (QLRO) ground state was recently investigated in a Heisenberg chain with interactions of the form 
$J_r \propto (-1)^{r-1}/r^\alpha$ \cite{laflorencie}. Here the signs correspond to no magnetic frustration, thus favoring AFM ordering. For 
$\alpha<\alpha_c$, the ground state possesses true AFM long-range order, while for $\alpha>\alpha_c$ the system is in a QLRO phase, with the same 
critical form of the spin correlations as in the standard Heisenberg chain. The critical value $\alpha_c$ depends on details of the couplings 
(e.g., on $J_1$ when all other $J_r$ are fixed) and the exponents are continuously varying. 

Another 
example of long-range interactions is the celebrated Haldane-Shastry chain \cite{haldane2}, with frustrated interactions $J_r = 1/r^2$. This system 
has a critical ground state similar to that of the standard Heisenberg chain, but the marginal operator vanishes \cite{essler} and it is, thus, 
a system right at the dimerization transition.

A natural question arising from previous work is how the combined effects of frustration and long-range interactions could lead to other phases 
and quantum phase transitions. In particular, is it possible to realize a direct transition between the AFM state and a VBS? In this  {\it Letter}
the evolution of the standard dimerization transition into an AFM--VBS transition is explored by considering a frustrated J$_{\rm 1}$-J$_{\rm 2}$ chain with 
additional {\it non-frustrated} long-range interactions. The hamiltonian for a finite periodic chain with $N$ spins $S=1/2$ is given by
\begin{equation}
H = \sum_{r=1}^{N/2} J_r \sum_{i=1}^N {\bf S}_i \cdot {\bf S}_{i+r},
\label{ham}
\end{equation}
where the couplings are given by
\begin{equation}
J_2=g,~~~J_{r \not= 2} = \frac{(-1)^{r-1}}{r^{\alpha}}\left (1+ \sum_{r=3}^{N/2}\frac{1}{r^\alpha}\right)^{-1}.
\end{equation}
Here the normalization is chosen such that the sum of all non-frustrated interactions $|J_r|$ equals $1$ \cite{jnote} (and $J_1$ is also given by
the $J_{r\not=2}$ expression). 

\begin{figure}
\centerline{\includegraphics[width=6.5cm, clip]{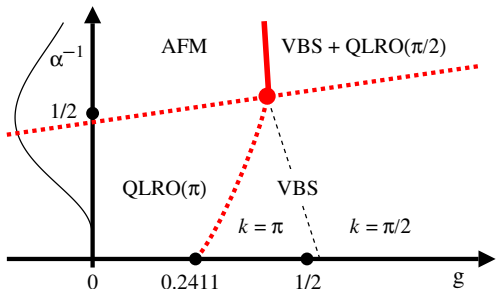}}
\vskip-2mm
\caption{(Color online) 
Approximate phase diagram as a function of the frustration strength $g$ and the inverse of the long-range 
exponent $\alpha$. The dashed curves indicate continuous transitions, whereas the thick solid curve represents 
a first-order transition. The curve for $g<0$ corresponds to the interaction studied in \cite{laflorencie}. 
At $\alpha^{-1}=0$ the dominant spin correlations in the VBS state change from $k=\pi$ to $\pi/2$  at 
$g\approx 0.52$ \cite{bursill}. This transition (or cross-over) evolves to the point where all phase 
boundaries come together.}
\label{fig1}
\vskip-3mm
\end{figure}

The model is here studied using Lanczos exact diagonalization. A semi-quantitative phase diagram based on these calculations in the plane $(g,\alpha^{-1})$ 
is shown in Fig.~\ref{fig1}. The J$_{\rm 1}$-J$_{\rm 2}$ chain corresponds to the horizontal axis ($\alpha^{-1}=0$). The QLRO phase is here denoted 
QLRO$(\pi)$, with $\pi$ indicating the wave-number of the dominant spin correlations. The phase boundaries are approximate, resulting primarily from
studies of level crossings, as will be discussed below. The main focus of this initial study of the model will be on the evolution of the QLRO$(\pi)$--VBS 
(dimerization) transition with decreasing $\alpha$. It will be shown that this continuous transition persists until $\alpha \approx 2$, while for smaller 
$\alpha$ it evolves into a first-order transition (of the avoided level-crossing type) between the AFM state and a state with coexisting VBS order and 
critical spin correlations at wave-number $k=\pi/2$, denoted in the phase diagram as VBS+QLRO$(\pi/2)$. 

The coexistence state is not purely of theoretical interest. Recent {\it ab-initio} calculations for metallic chains show unfrustrated spin couplings 
decaying as $\approx 1/r^2$,  with $J_2$ in some cases frustrating (e.g., Mn) \cite{tung}. In the quasi-classical (large-$S$) limit, spiral states with 
continuously varying periodicity can arise in such a systems. The present study suggests a more exotic scenario in the extreme quantum limit of $S=1/2$ 
(and perhaps also for other small $S$).

`Solving the model (1) numerically poses significant technical challenges. Efficient quantum Monte Carlo 
techniques can be applied to systems with long-range interactions \cite{sandvik1,laflorencie}, but with the frustrating $J_2$ term this is no longer 
possible due to the sign problem \cite{henelius}. The DMRG method \cite{white,schollwock}, on the other hand, can handle frustration but not easily 
long-range interactions. Here periodic chains up to size $N=32$ are solved using Lanczos exact diagonalization (in the standard way, exploiting lattice 
symmetries and spin-inversion for block-diagonalization in the magnetization $m_z=0$ sector). This is sufficient for roughly extracting the phase 
boundaries using level crossing methods (which in the case of the dimerization transition is a well established technique \cite{nomura}, extended 
here using different levels to detect other transitions).

\begin{figure}
\centerline{\includegraphics[width=7cm, clip]{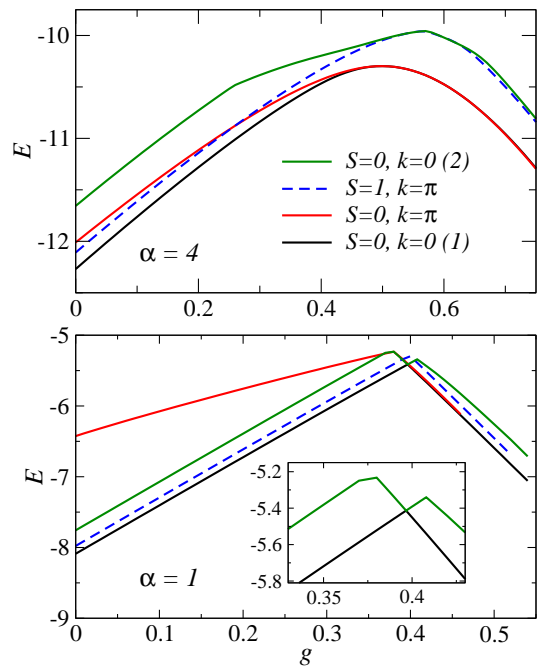}}
\vskip-2mm
\caption{(Color online) 
Low-energy levels of a $16$-spin system at  $\alpha=4$ (upper panel) and $\alpha=1$ (lower panel).
The spin $S$ and the momentum $k$ of the states are indicated in the upper panel. The inset in the lower panel 
shows the avoided level crossing of the two $k=0$ singlets in greater detail.}
\label{fig2}
\vskip-3mm
\end{figure}

\begin{figure}
\centerline{\includegraphics[width=7.5cm, clip]{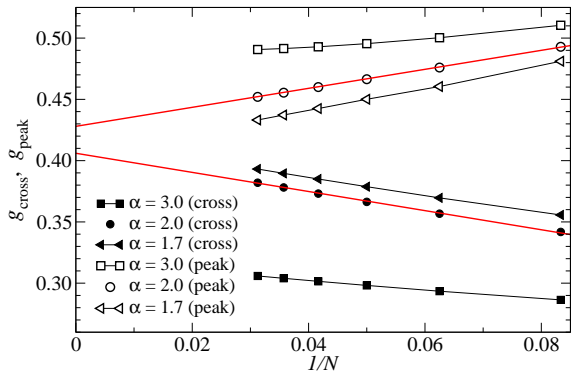}}
\vskip-2mm
\caption{(Color online) Dependence on the inverse chain length of the singlet-triplet crossing point $g_{\rm cross}$ and the location
$g_{\rm peak}$ of the ground state energy maximum for different long-range interaction exponents $\alpha$. The two lines show extrapolations 
of the $g=2.0$ numerical data to $N=\infty$.}
\label{fig3}
\vskip-3mm
\end{figure}

The QLRO$(\pi)$--VBS transition in the J$_{\rm 1}$-J$_{\rm 2}$ chain is of infinite order, i.e., the singlet-triplet gap of the VBS is exponentially 
small for $g \to g_c$ \cite{affleck1}. It is therefore difficult to locate the transition based on the order parameter for small $N$. However, 
$g_c$ can be determined accurately from excited states. The lowest excitation of a chain with even $N$ is a triplet for $g<g_c$ and a singlet 
for $g>g_c$. The crossing point of these levels is a rapidly converging finite-$N$ definition of $g_c$ \cite{nomura,eggert}. 
The same physics can be expected also in the presence of the long-range interaction, if $\alpha$ is sufficiently 
large. This is shown for a  $16$-spin chain at $\alpha=4$ in the upper panel of Fig.~\ref{fig2}. Singlets with momenta $k=0$ and $k=\pi$ (out of 
which symmetry-broken dimerized states can be formed) should be degenerate in the VBS phase. For finite $N$ this degeneracy is not exact (except in the 
J$_{\rm 1}$-J$_{\rm 2}$ chain at the special point $g=1/2$), but a region of very near degeneracy for $g>1/2$ can be seen in the figure. The region 
of approximate degeneracy, which is not easy to demarcate precisely, expands very slowly toward smaller $g$ with increasing $N$. In contrast, the
singlet-triplet crossing point is well defined and converges rapidly. Extrapolating the crossing point to $N=\infty$ for different $\alpha$, as
illustrated in Fig.~\ref{fig3}, can reliably give the QLRO$(\pi)$--VBS phase boundary $g_c(\alpha)$ for $\alpha \agt 2$.

Upon decreasing $\alpha$ below $\approx 2$, the broad maximum in the ground state energy versus $g$  becomes increasingly sharp. As 
seen in the lower panel of Fig.~\ref{fig2}, at $\alpha=1$ it has developed into a sharp tip due to an avoided level crossing with the second singlet at 
$k=0$. The real singlet-triplet crossing has moved to the same region. An avoided level crossing leading to a discontinuity in the derivative of the 
ground state energy with respect to $g$ for $N\to \infty$ is the hall-mark of a first-order transition. The nature of the phases at this transition 
will be discussed below. First, let us investigate how the transition evolves from continuous to first-order.

Fig.~\ref{fig3} shows the size dependence of the level crossing point $g_{\rm cross}$ and the location $g_{\rm peak}$ of the maximum in the 
ground state energy. In the J$_{\rm 1}$-J$_{\rm 2}$ chain the size correction to the crossing point is $\propto 1/N^2$, which also can be seen
for large $\alpha$. For smaller $\alpha$, the corrections instead seem to be  $\propto 1/N$, but a cross-over to $1/N^2$ for large $N$ seems 
likely as long as the transition remains continuous.  The peak location moves in the opposite direction. For some $\alpha$ and $N\to \infty$,
$g_{\rm cross}$ and $g_{\rm peak}$ should coincide. The results indicate that both $g_{\rm cross}$ and $g_{\rm peak}$ have dominant $1/N$ corrections 
at this point. Fitted lines are shown in Fig.~\ref{fig3} at $\alpha=2$, were there is still a small gap between the two extrapolated values. 
For $\alpha=1.7$, where the transition is first-order, they should coincide (and then the asymptotic size correction should be exponential).

\begin{figure}
\centerline{\includegraphics[width=7.4cm, clip]{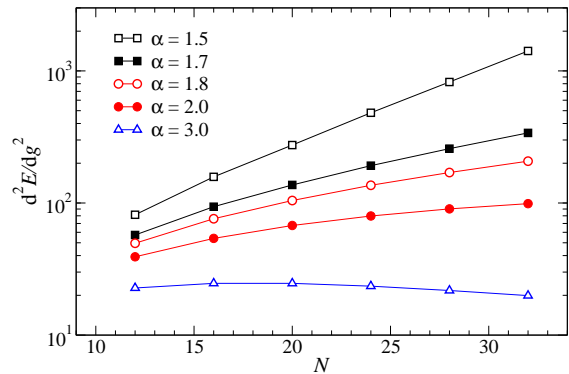}}
\vskip-2mm
\caption{(Color online) Size dependence of the second derivative of the ground state energy with respect to the frustration parameter $g$ at the 
point $g_{\rm peak}$ where the ground state energy takes its maximum value.}
\label{fig4}
\vskip-3mm
\end{figure}

To verify an avoided level crossing with a discontinuous energy derivative for $\alpha \alt 1.8$, the second derivative of the ground state
energy at its maximum is graphed on a lin-log scale in Fig.~\ref{fig4}. It grows exponentially with $N$ for $\alpha=1.5$, showing that the slope of 
the energy curve indeed changes discontinuously for an infinite chain. In contrast, at $\alpha=3$ the second derivative decreases for large $N$. 
For $\alpha=2$ convergence to a finite value also seems plausible, whereas $\alpha=1.7$ and $1.8$ appear to be close to a separatrix 
(where the form of the divergence is consistent with a power law) between the two different behaviors. 

This analysis suggests that the continuous dimerization transition changes smoothly into a first-order transition at 
$(g_m \approx 0.41,\alpha_m \approx 1.8)$. The singlet-triplet crossing moves toward the ground state energy maximum and coincides with it 
at the multi-critical point $(g_m,\alpha_m)$, beyond which it develops into a first-order singularity. Note that the rounded energy maximum
in the VBS phase for large $\alpha$ has no special significance (except at $\alpha^{-1}=0$ where it corresponds to the exact singlet-product
ground state). It is only when it develops into the sharp avoided level crossing that it is associated with a phase transition.

\begin{figure}
\centerline{\includegraphics[width=6.25cm, clip]{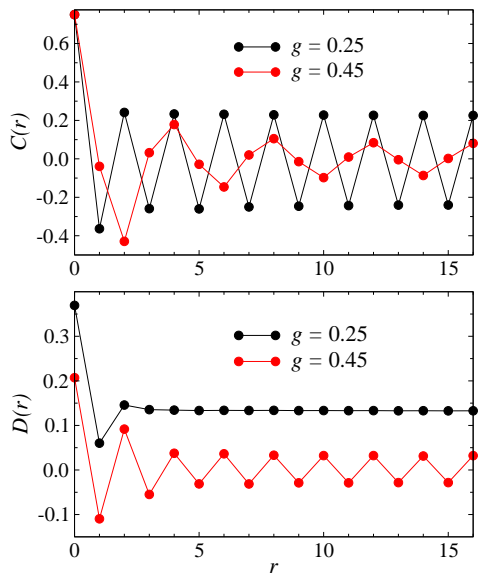}}
\vskip-2mm
\caption{(Color online) 
Spin (upper panel) and dimer (lower panel) correlations in a 32-spin chain at $\alpha=1$.
At $g=0.25$ and $0.45$ the system is in the AFM and VBS--QLRO$(\pi/2)$ phase, respectively. 
A first-order transition between these states occurs at $g\approx 0.39$.}
\label{fig5}
\vskip-3mm
\end{figure}

To discuss the states involved in the first-order transition, consider the spin and bond correlation functions;
\begin{eqnarray}
C(r) & = & \langle {\bf S}_i \cdot {\bf S}_{i+r}\rangle ,\\
D(r) & = & \langle ({\bf S}_i \cdot {\bf S}_{i+1})({\bf S}_{i+r} \cdot {\bf S}_{i+r+1})\rangle .
\end{eqnarray}
In Fig.~\ref{fig5} these are graphed for two $g$ values, at either side of the transition for $\alpha=1$. At $g<g_c$ the dominant spin correlation 
$C(k)$ in Fourier space is at wave-number $k=\pi$, and finite-size scaling shows that the sublattice magnetization remains non-zero for 
$N\to \infty$. There is no structure in $D(r)$, i.e., there is no VBS order. This is thus an AFM phase; the continuation of the AFM state studied 
in \cite{laflorencie}, as indicated in Fig.~\ref{fig1}. For $g>g_c$ there is VBS order. Interestingly, in this phase there are also strong spin 
correlations at $k=\pi/2$, which can be seen clearly as a real space period-four oscillation in Fig.~\ref{fig5}. Finite-size scaling indicates 
that there is no long-range spin order, but the correlations appear to decay as $1/r^\gamma$ with $\gamma\approx 1$; thus this state is denoted as
QLRO$(\pi/2)$. 

Examining the correlations as a function of $g$, discontinuities (increasingly sharp jumps with increasing $N$) develop for $\alpha < 1.5$. This 
should persist until the multi-critical point at $\alpha_m\approx 1.8$, but larger systems are needed to observe the discontinuity very
close to this point. 

The VBS+QLRO$(\pi/2)$ state should have gapless spin excitations. The lowest triplet has $k=\pi/2$. It is, however, difficult to 
demonstrate the gaplessness based on data for small systems, because the size-dependence of the gaps (and other quantities) for $N=4n$ exhibit 
even-odd oscillations in $n$. In the VBS phase the lowest triplet is at $k=\pi$, even when the spin correlations (exponentially decaying)
are peaked at $k=\pi/2$. The level crossing between the lowest $k=\pi$ and $k=\pi/2$ triplets can be used to extract the boundary between the 
VBS and VBS+QLRO$(\pi/2)$ phases. The size dependence of the crossing point is not smooth, however, and cannot be extrapolated very reliably. 
The boundary between dominant $k=\pi$ and $k=\pi/2$ spin correlations in the VBS phase has also not been extracted accurately. 

Let us briefly return to Fig.~\ref{fig2} for another interesting feature of the level spectrum: The lowest singlet excitation for small $g$ has 
momentum $k=\pi$ for $\alpha=4$ but $k=0$ for $\alpha=1$. The switching of the order of these levels as a function of $\alpha$ for $g<g_c$ is 
associated with the QLRO$(\pi)$-AFM transition. The level crossings can be used to extract this phase boundary very accurately up to 
$g \approx 0.25$ (while for higher $g$ the $N\to \infty$ extrapolations become difficult); a more detailed discussion of this issue is given
as a footnote \cite{crossnote}. As indicated in Fig.~\ref{fig1}, $\alpha_c$ depends only weakly on $g$. The results are consistent with the 
location quoted above for the multi-critical point.

In summary, the combination of short-range frustration and long-range unfrustrated interactions in one dimension has been shown to lead
to a first-order transition between a N\'eel state and a VBS with coexisting critical $k=\pi/2$ spin correlations. It should be noted that
the system sizes studied here are small, and it cannot be excluded that the spin correlations could become incommensurate, as they do in
the J$_{\rm 1}$-J$_{\rm 2}$ chain for $J_2/J_1>1$.\cite{kumar} Hopefully, field-theories that very successfully describe the standard dimerization 
transition \cite{affleck1}, and recently also the transition between the critical spin state and the N\'eel state \cite{laflorencie}, could 
be generalized to the coexistence state as well. 

Recent calculations \cite{tung} for metallic chains have shown that interactions of the type used here are realistic, but in these systems $S>1/2$.
Although one cannot describe these systems completely using a spin-only model, it would still be interesting to repeat the calculations reported
here for larger $S$. This is much more challenging, however, because of the rapidly growing size
of the Hilbert space with $S$. Although the DMRG method \cite{white,schollwock} is not ideally suited for systems with long-range interactions, 
it may still be possible to use it to study lattice sizes beyond the limits of Lanczos calculations.

{\it Acknowledgments}---I thank I. Affleck and G. Y. Guo for stimulating discussions, and G. Y. Guo also for communicating
the results of Ref.~\cite{tung} before publication. This work is supported by NSF grant No.~DMR-0803510. 

\null


\begin{thebibliography}{00}
\bibitem{jnote}
This guarantees a finite energy per spin for $N\to \infty$ even for $\alpha<1$. Instead of summing $J_r$ up to $r=N/2$, one could also include 
$N/2 < r < N$. This should not affect the phase boundaries and critical exponents for $\alpha>1$.
\bibitem{crossnote}
In the QLRO$(\pi)$ phase the low-energy excitations arise from two deconfined spinons. The lowest (for $N=4n$) is a triplet 
at $k=\pi$. The lowest singlet is also at $k=\pi$, with a small finite-size gap to the triplet due to weak spinon-spinon interactions. In the 
AFM phase the spinons are no longer deconfined and the $k=\pi$ singlet energy is higher. The Lanczos calculations show that the lowest singlet 
is at $k=0$. In \cite{laflorencie} the QLRO$(\pi)$--AFM transition in a model with $J_1=1$ and $J_{r>1}=\lambda (-1)^{r-1}/r^\alpha$ was located 
using quantum Monte Carlo data for the order parameter, which gave $\alpha_c=2.225 \pm 0.025$ at $\lambda=1$ using $N$ up to $4096$. Extrapolating 
the $k=0,\pi$ singlet crossing points (which have size corrections $\propto 1/N^{\beta}$, with $\beta\approx 1.50$) for $N\le 32$ gives a marginally 
higher value; $\alpha_c=2.262 \pm 0.001$. Analyzing the singlet and triplet gaps at the crossings, assuming $\Delta \sim N^{-z}$, gives the dynamic 
exponent $z=0.764 \pm 0.005$, in very good agreement with \cite{laflorencie}. For the frustrated model (\ref{ham}) on the QLRO$(\pi)$--AFM boundary 
$z\approx 0.75$ for $g$ up $\approx 0.25$, while for larger $g$ it is not possible to reliably extract $\alpha_c$ and $z$ this way, because of 
large scaling corrections and the absence of level crossings for increasingly large systems as the multi-critical point 
$(g_m\approx 0.41,\alpha_m \approx 1.8)$ is approached (for reference $\alpha_c = 2.220 \pm 0.005$ for $g=0$ and $2.170 \pm 0.01$ for $g=0.2$). 
At the multi-critical point $z$ can be extracted using gaps at the singlet-triplet crossing point and the ground state energy maximum, giving
$z\approx 0.8$.
\end{thebibliography}
\end{document}